\documentclass[10pt,journal,compsoc]{IEEEtran}
\pdfoutput=1

\usepackage[T1]{fontenc}
\usepackage[utf8]{inputenc}

\usepackage[usenames, dvipsnames]{color}
\usepackage{soul}

\usepackage[hyphens]{url}
\usepackage{hyperref}

\usepackage{multirow}
\usepackage{booktabs}

\usepackage{amsmath,amssymb}
\DeclareMathOperator{\E}{\mathbb{E}}

\usepackage{graphicx}

\renewcommand\paragraph[1]{\medskip\noindent\textbf{#1:}}

\makeatletter
\def\@oddfoot{\mycopyrightnotice}
\def\ps@IEEEtitlepagestyle{
  \def\@oddfoot{\mycopyrightnotice}
}
\def\mycopyrightnotice{
  {\sffamily\scriptsize
  \copyright 2022 IEEE. Personal use is permitted, but republication/redistribution requires IEEE permission. See https://www.ieee.org/publications/rights/index.html for more information.
  }
}

\begin{document}

\title{A Case for Partitioned Bloom Filters}

\author{Paulo S\'ergio Almeida \\ INESC TEC and University of Minho}
\date{}

\IEEEtitleabstractindextext{%
\begin{abstract}
In a partitioned Bloom Filter (PBF) the bit vector is split into disjoint parts,
one per hash function. Contrary to hardware designs, where they prevail,
software implementations mostly ignore PBFs, considering them worse than
standard Bloom filters (SBF), due to the slightly larger false positive rate
(FPR). In this paper, by performing an in-depth analysis, first we show that
the FPR advantage of SBFs is smaller than thought; more importantly,
by deriving the per-element FPR, we show that SBFs have weak spots in the
domain: elements that test as false positives much more frequently than expected.
This is relevant in scenarios where an element is tested against many filters.
Moreover, SBFs are prone to exhibit extremely weak spots if naive double
hashing is used, something occurring in mainstream libraries. PBFs exhibit a
uniform distribution of the FPR over the domain, with no weak spots, even
using naive double hashing. Finally, we survey scenarios beyond set membership
testing, identifying many advantages of having disjoint parts, in designs
using SIMD techniques, for filter size reduction, test of set disjointness,
and duplicate detection in streams. PBFs are better, and should replace SBFs,
in general purpose libraries and as the base for novel designs.
\end{abstract}

\begin{IEEEkeywords}
  Probabilistic data structures, Information filtering, Partitioned Bloom filters
\end{IEEEkeywords}}

\maketitle

\IEEEpeerreviewmaketitle

\section{Introduction}

A Bloom filter~\cite{Bloom70} is a probabilistic data structure to represent
a set in a compact way. An element which has been inserted will always be
reported as present; an element not in the set may erroneously be reported
as present (i.e., false positives may arise), but the Bloom filter may be
configured such that the probability of false positives may be as low as
desired.
Bloom filters are used in many settings, such as
networking~\cite{broder2004network} and distributed systems~\cite{tarkoma2012theory}.

A standard Bloom filter is a single array of $m$ bits over which $k$
independent hash functions range. When inserting an element, each of the $k$
functions is used to produce an index, and the corresponding bit is set. When
querying, an element is considered present if all bits in the positions given
by the $k$ hash functions are set.

A variant, partitioned Bloom filters, proposed by
Mullin~\cite{Mullin1983}, divides the array into $k$ disjoint parts of size $m/k$
(assuming $m$ multiple of $k$). Each of the $k$ hash functions ranges over
$m/k$, being used to set or test a bit in the corresponding part.  The more
obvious feature in partitioned Bloom filters is the complete independence of
each of the $k$ parts and of each corresponding bit setting/testing. This
has some obvious advantages, such as parallel access to each part, which has
made partitioned Bloom filters widely adopted in hardware
implementations~\cite{ceze2006bulk,sanchez2007implementing}, where they are
sometimes called \emph{parallel Bloom signatures}.

A hybrid variant divides the filter in $k/h$ parts, with $h$ hash functions
per part, such as a hardware implementation~\cite{dharmapurikar2003deep}
where $k/h$ independent multi-port memory cores, each allowing $h$ accesses
per cycle is used. An important consideration~\cite{sanchez2007implementing}
for hardware designs is that using single-port SRAM,
for the partitioned scheme, requires much less area than using $k$-ported SRAM
for the standard scheme, or $h$-ported SRAM for the hybrid scheme, because the size
of an SRAM cell increases quadratically with the number of ports. This seems
to settle the standard-versus-partitioned choice for hardware designs, leading
them to typically opt for the partitioned variant.

Concerning software implementations, standard Bloom filters prevail. The
general feeling towards partitioned Bloom filters is that they are almost the
same as standard ones, but produce slightly worse false positive rate (FPR),
specially in small Bloom filters. This comes from the
observation~\cite{KM2008} that partitioned Bloom filters will have slightly
more bits set than standard ones, and this slightly higher fill ratio
(proportion of set bits) will result in a correspondingly higher FPR.

As we will demonstrate in this paper, the issue is more subtle, and
this slight advantage comes at a substantial cost, including in the false
positive rate itself. The main contributions of this paper are:

\begin{itemize}
    \item Perform an in-depth analysis of the FPR in Bloom
      filters where we: provide a simpler explanation, compared with current
      literature, of why the standard formula is a strict lower bound of the
      true FPR; address the effect due to different hash
      functions colliding for a given element; obtain for the first time an
      exact formula for the \emph{per-element} FPR, i.e., the
      expected FPR, for each specific element of the domain, over
      the range of filters that do not contain it.
    \item Point out the consequences for standard Bloom filters of the
      above hash collision problem, namely the occurrence of \emph{weak spots} in
      the domain: elements which will be tested as false positives much more
      frequently than expected. This can be a problem, both for
      standard small capacity Bloom filters, or for \emph{blocked Bloom
      filters}~\cite{PutzeSS09}, and its unexpectedly frequent occurrence can be
      as surprising as the \emph{Birthday Problem}~\cite{MRE1939}.
    \item Expose pitfalls when using \emph{Double Hashing} with standard Bloom
      filters, of which many widespread libraries seem to be unaware off,
      and contrast it with the robustness of partitioned Bloom filters
      in this matter.
    \item Survey usages for Bloom filters other than testing set membership,
      identifying many advantages that result from having disjoint parts that
      can be individually sampled, extracted, added or retired. We identify
      how the partitioned scheme leads to superior designs for SIMD
      techniques, testing set disjointness, reducing filter size, and
      duplicate detection in streams.
\end{itemize}

\section{Bloom filters and the Birthday Problem}

\begin{figure}
  \begin{center}
    \includegraphics{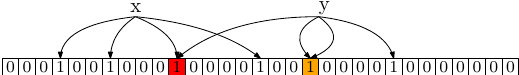}
  \end{center}
\caption{Standard Bloom filter using 4 hash functions.}
\label{fig:sbf1}
\end{figure}

While most Bloom filters are used to represent large sets, in some scenarios
small Bloom filters are used. If a small FPR is also wanted,
the combination of a small $m$ and a (relatively) large $k$ will cause, for a
standard Bloom filter, a non-negligible probability that two or more of the $k$ hash
functions, applied to a given element, collide (produce the same index).
Such a collision is illustrated in Figure~\ref{fig:sbf1}, in yellow, where two of the 4
hash functions applied to $y$ produce the same index, resulting in a total of
three bits being set for $y$, instead of the expected 4 bits.
Such intra-element hash collisions are not normally illustrated (or discussed)
in Bloom filter presentations, which just focus on inter-element collisions, such as
the one between $x$ and $y$, in red. 

In fact the surprisingly high probability of intra-element hash collisions is
precisely an instance of the \emph{Birthday Problem}, stated in 1927 by H.
Davenport\footnote{But frequently misattributed to von Mises, who stated a
similar but different version of the problem. Some archaeology about its
origin can be found at~\cite{patblog2011}.}, as described by Coxeter~\cite{MRE1939}.
The probability that, for a given element, two or more of the $k$ independent
hash functions return the same value is:
\begin{equation} \label{eq:birthday}
1 - {P(m, k) \over  m^k }, \end{equation}
where $P(m,k)$ denotes the $k$-permutations of $m$. We now give some examples.

\paragraph{Sets of words in small strings}
Mullin~\cite{mullin1987accessing} used Bloom filters to store sets of
words occurring in strings (e.g., titles and authors of articles), typically
up to 15 words per string, with filters ranging from 32 up to 256 bits,
the most common one being 96 bits, and using 8 hash functions per filter.
With $m=96$ and $k=8$ two or more hash function will collide in one out of
four cases (25.88\%), where the FPR will be at least twice
the expected from the classic formula (for filters that reached design capacity),
or much higher than expected (for filters still far away from design capacity).

\paragraph{Packet forwarding}
Whitaker and Wetherall~\cite{whitaker2002forwarding} used small Bloom filters
in packets to detect possible forwarding loops in experimental routing protocols.
In this case 64 bit filters were used, with ``4 bits set to one''.
With $m=64$ and $k=4$ two or more hash function will collide 9.1 percent of
the time. Interestingly, and different from the more normal usage, in this
case a given element (node) is tested against many Bloom filters (packets),
and instead of using $k$ hash functions for the element, a \emph{Bloom mask}
with exactly 4 ones at random positions is computed at start time,
overcoming the collision problem.

\paragraph{Blocked Bloom filters}
One problem with Bloom filters is the spreading of memory accesses, hurting
performance. This is avoided by \emph{blocked Bloom filters}~\cite{PutzeSS09},
where the filter is divided into many blocks, each block a Bloom filter
fitting into a single cache line (e.g., 512 bits), and using an extra hash
function to select the block. For a very high precision filter, with $k=16$ and
$m=512$, hash collisions will occur for 21 percent of elements, and even for a
more normal setting of $k=8$, there will be collisions for 5.3 percent of
elements. For an extreme performance BBF that requires a single memory access,
using word sized blocks, $m=64$, for $k=8$ we have collisions 36 percent of
time, or 9 percent of the time for the more reasonable $k=4$. So, the
collision problem occurs in practice for BBFs. It should be emphasized that
using blocking is the only way that Bloom filters can remain performance-wise
competitive~\cite{LangNKB19} with dictionary-based approaches (such as Cuckoo
Filters~\cite{FanAKM14}, Morton Filters~\cite{BreslowJ18}), or Xor
Filters~\cite{GrafL20}. Therefore, the
scenario of a small Bloom filter (a block of a BBF) is important, even for
scenarios with huge (on the whole) filters.

\medskip

\begin{figure}
  \begin{center}
    \includegraphics{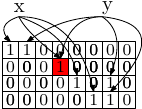}
  \end{center}
\caption{Partitioned Bloom filter using 4 hash functions, represented as a
bidimensional bit array with one row per part.}
\label{fig:pbf1}
\end{figure}

The above mentioned hash collision possibility is not a problem in partitioned
Bloom filters because each of the $k$ functions is used to set/test bits in a
different part.  While in standard Bloom filters hash collisions will lead to
bit collisions (the same bit being used for different functions), in
partitioned Bloom filters such hash collisions will not lead to bit
collisions. This is illustrated in Figure~\ref{fig:pbf1}, which shows a
partitioned Bloom filter using 4 parts, represented as a bidimensional bit
array with one row per part. It can be seen that even if two of the 4 hash
functions applied to $y$ produce the same value (column index), two different
bits in the filter are set.

So, while for partitioned Bloom filters, \emph{exactly} $k$ distinct bits in
the filter are accessed, in standard Bloom filters \emph{up to} $k$ distinct
bits are accessed (most times $k$ bits, but sometimes less than $k$ bits).  As
we will see, this makes the standard false positive formula incorrect,
producing a value lower than the actual one, and complicating the exact false
positive calculation (something that has been addressed before) but it also
produces a \emph{non-uniform distribution} of the FPR, with the
occurrence of \emph{weak spots} in the domain, something that we address here
for the first time.

Interestingly, in the original proposal by Bloom~\cite{Bloom70} exactly $k$
bits are set/tested: ``each message in the set to be stored is hash coded into
a number of distinct bit addresses'' and ``where d is the number of distinct
bits set to 1 for each message in the given set''. The original formula for
the FPR is consistent with this behavior.  This fact seems to have been mostly
ignored in the literature, being one notable
exception~\cite{manber1994algorithm} ``In [Bl70], the assumption was that the
k locations are chosen without repetitions; it is also possible to allow
repetitions, which makes the program simpler'' and more
recently~\cite{Grandi2018} a comparison between the original proposal and standard
Bloom filters.

The original Bloom proposal is not practical, as it demands some extra effort
to ensure exactly $k$ distinct addresses, e.g., iterating over an
unbounded family of hash functions until $k$ different values have been
produced (with the need to compare each new value to all the previous ones);
or a way to directly produce a pseudo-random $k$-permutation of $m$,
keyed by the element. And even if little cost seems to be
required~\cite{Roberts79}, practitioners typically would not be aware of the
problem or solution, and would not bother to address such minutiae. So, it is
not surprising that what became adopted as standard Bloom
filters differs from the original proposal.

Partitioned Bloom filters, which differ both from the original and the standard
ones, not only are immune to the birthday problem (being in a sense more in
the spirit of the original proposal) but are also practical to implement.

\section{False positive analysis}

We now do a theoretical analysis of the FPR, revisiting the
Bloom's analysis, the standard analysis, existing improvements to the standard
analysis producing a correct formula, the formula for partitioned Bloom
filters, and compare standard with partitioned Bloom filters. In the next
section we present a novel \emph{per-element} false positive analysis, showing
how the expected FPR behaves for different elements in the domain.

\subsection{Original Bloom's analysis}

Bloom's analysis~\cite{Bloom70} states that the probability of a bit still
being zero after $n$ elements are added is
\begin{equation}
  \left(1 - {k \over m}\right)^n,
\end{equation}
which, contrary to what sometimes is said, is correct, but for the original
Bloom proposal where exactly $k$ \emph{distinct} bits are set, and that the false
positive rate is:
\begin{equation}
  \left(1 - \left(1 - {k \over m}\right)^n\right)^k.
\end{equation}
The analysis is almost correct, but it suffers from the same problem as the
standard analysis below. But it is irrelevant for standard Bloom filters used
in practice, as they differ from the original Bloom proposal.

\subsection{Standard analysis}

The standard analysis, by Mullin~\cite{Mullin1983}, and widely used,
states that the probability of a bit still being zero after $n$
elements are added is
\begin{equation}
  \left(1 - {1 \over m}\right)^{kn},
\end{equation}
which is correct, and that the FPR is
\begin{equation}
  \label{eq:standard}
  F_a(n, m, k) = \left(1 - \left(1 - {1 \over m}\right)^{kn}\right)^k.
\end{equation}
which is only approximate, as we discuss below.

\subsection{The exact formula for standard Bloom filters}

There is one problem with the standard analysis, which has already been detected
and corrected before. The standard analysis derives the FPR only as
function of the mean fill ratio $p$, as $p^k$. Even though this gives a very
good approximation for large Bloom filters, given the high concentration of
the fill ratio around its mean~\cite{mitzenmacher2002compressed}, it is not an
exact formula.

Exact formulas for standard Bloom filters were
developed~\cite{Bose2008210,christensen2010new}, by
deriving the probability distribution of the fill ratio and weighing the false
positive rate incurred by each concrete fill ratio with the probability of it
occurring. A similar result had already been
derived~\cite{manber1994algorithm}, for a Bloom filter variant divided in
pages (essentially, a blocked Bloom filter with typically large blocks), and a
formula for the original Bloom filters was derived more recently~\cite{Grandi2018}.

\paragraph{A simpler strict lower bound argument}
The standard formula, in Equation~\ref{eq:standard}, has also been proven to be
a strict lower bound for the true FPR~\cite{Bose2008210}
using considerations of conditional probability, and to be a lower
bound~\cite{christensen2010new} by resorting to H\"older's
inequality~\cite{Holder1889}.
We now present a simpler and more elegant reasoning of why it
is a strict lower bound. It results from a direct application of Jensen's
inequality~\cite{jensen1906}: for a convex function, such as $f(x) = x^k$ when
$k > 1$ and $x > 0$, and for a non-constant random variable $R$, such as the
fill ratio,
\begin{equation}
  \label{eq:jensen}
  f(\E[R]) < \E[f(R)].
\end{equation}
This means that, for $k > 1$, raising the expected fill ratio to the
power of $k$, as done in the standard formula, produces a value always smaller
than the expected value of the fill ratio raised to the power of $k$, which is
what gives the exact average FPR.

\medskip

As presented by the above mentioned works, computing the fill ratio
distribution is an instance of the well known \emph{balls into bins}
experiment.
It can be
computed by resorting to the number of surjective functions from an $n$-set to
an $i$-set, $e_{ni}$~\cite{gerrish1979}, that can be directly derived using
the inclusion-exclusion principle (in the complementary form) as:

\begin{equation}
  e_{ni} = \sum_{j = 0}^{i} (-1)^j {i \choose j} (i - j)^n.
\end{equation}

The probability $B(n, m, i)$ of having exactly $i$ non-empty bins, after
throwing $n$ balls randomly into $m$ bins is then:

\begin{equation}
  \label{eq:balls-bins}
  B(n, m, i) = {{m \choose i} e_{ni} \over m^n}.
\end{equation}

The probability of having exactly $i$ bits set after inserting $n$ elements
into an $m$ sized standard Bloom filter using $k$ hash functions is then:
\begin{equation}
  S(n, m, k, i) = B(nk, m, i) ,
\end{equation}
and the FPR for a standard Bloom filter is then:

\begin{equation}
  \label{eq:fps}
  F_s(n, m, k) = \sum_{i=1}^m S(n, m, k, i) \left({i \over m}\right)^k .
\end{equation}

\subsection{The exact formula for partitioned Bloom filters}

As the $k$ parts are independently set/tested, the expected FPR is the product
of the individual expected rates, and so computed as the one for each part to
the power of $k$.  For each part, the standard formula, with $k=1$, gives the
exact part FPR, as the inequality in Equation~\ref{eq:jensen} becomes an
equality when $k=1$.  So, for a partitioned Bloom filter of size $m$, made up
of $k$ parts, each $m/k$ bits, the exact FPR when $n$ elements were inserted
is given by:
\begin{equation}
  F_p(n, m, k) = \left(1 - \left(1 - {k \over m}\right)^n\right)^k ,
\end{equation}
which is much simpler than the exact formula for standard Bloom filters (as
well as the exact formula for original Bloom filters~\cite{Grandi2018}).
Interestingly, it coincides with Bloom's formula for his original proposal,
while being exact.

This formula simplicity results from the conceptual simplicity: a partitioned
Bloom filter can be seen as an AND of $k$ independent single-hash filters, all
used for each insertion. It also translates to a simplicity of presentation, which
is better, pedagogically, than standard Bloom filters, as it allows deriving a
more complex (composite) concept in terms of a simpler one (single-hash filters).

\subsection{Comparison with partitioned Bloom filters}

Common folklore is that partitioned Bloom filters are not worth over standard
ones~\cite{KM2008}: ``partitioned filters tend to have more 1's than
nonpartitioned filters, resulting in larger false positive probabilities''.
But in spite of hash collisions decreasing the fill ratio, they increase
the false positives for elements with collisions, and so the question
is more subtle. Using the exact formulas for each case,
Table~\ref{tab:std-part-fp-comparison} shows how partitioned and standard Bloom
filters compare, namely the ratio of false positives $F_p / F_s$, for some
combinations of $m$ and $k$ for filters at full capacity with $n = \frac mk
\ln 2$.

We focus on small/medium sized filters for two reasons: 1) for large
(plain) filters standard and partitioned variants exhibit almost the same
FPR, being essentially indistinguishable; there is no point in comparing them
and; 2) even if we want a large (on the whole) filter, the best performance
will be achieved using a blocked Bloom filter, which will have relatively
small blocks.
The three sizes chosen (64, 512, and 4096 bits) are the more important ones for
which it is meaningful to compare SBF and PBF. The first two, the word sized
(64 bits) filter and the cache line sized (typically 512 bits) filter, the
typical block in a BBF, are the more important ones. The last one (4096 bits)
is still small enough such that some difference between SBFs and PBFs can be
observed. The trio forms a sequence, each pair roughly separated by (almost)
an order of magnitude (512/64 = 4096/512 = 8).

\begin{table}
  \caption{Comparison between partitioned and standard Bloom filters FPRs, for different combinations of $m$ and $k$, for filters at nominal
    occupation ($n = \frac mk \ln 2$), showing both the approximate ($F_a$) and
the exact ($F_s$) values for standard filters, the value for partitioned
filters ($F_p$) and the ratio $F_p / F_s$.}
  \label{tab:std-part-fp-comparison}
  \begin{center}
    \begin{tabular}{@{}rrcccc@{}}
  \toprule
  $m$ & $k$ & $F_a$ & $F_s$ & $F_p$ & $F_p / F_s$ \\
  \midrule
  \multirow2*{64}
  & 4 & 0.06244514 & 0.06423247 & 0.06676410 & 1.03941360 \\
  & 8 & 0.00227672 & 0.00260362 & 0.00316870 & 1.21703762 \\
  \midrule
  \multirow3*{512}
  & 4 & 0.06126247 & 0.06148344 & 0.06176528 & 1.00458411 \\
  & 8 & 0.00375309 & 0.00381650 & 0.00389940 & 1.02172097 \\
  & 16 & 0.00001409 & 0.00001513 & 0.00001661 & 1.09783475 \\
  \midrule
  \multirow3*{4096}
  & 4 & 0.06233016 & 0.06235819 & 0.06239353 & 1.00056676 \\
  & 8 & 0.00385474 & 0.00386284 & 0.00387308 & 1.00265094 \\
  & 16 & 0.00001486 & 0.00001499 & 0.00001516 & 1.01143019 \\
  \bottomrule
    \end{tabular}
  \end{center}
\end{table}

It can be seen that although partitioned filters have indeed slightly more
false positives, the difference is less than what the standard formula ($F_a$)
would suggest, and for all purposes irrelevant in practice. The largest
increase is for the word sized Bloom filter with $k=8$, with 22\% higher
FPR. This extra 22\%, which would be significant for
other metrics like CPU usage, is not very relevant for the FPR, where mostly
orders of magnitude matter. In this specific example, a PBF would have the
same FPR as a standard filter by reducing the nominal capacity to compensate,
roughly to $1 / (1.22^{(1/8)}) = 0.975$, so by having only 2.5\% less
capacity than a standard BF. Moreover this combination $m=64$ and $k=8$ is an
extreme case, more for illustration purposes, as it is not really suitable,
allowing just a few elements in the filter or, in the case of a BBF, would
lead to FPR degradation due to the danger of some blocks being too
overloaded~\cite{PutzeSS09} (occupancies of blocks of a BBF follow a binomial
distribution). BBFs normally aim for larger blocks of cache line size,
typically with $m=512$, with word sized blocks not causing significant FPR
degradation only for smaller values of $k$ (and larger FPRs).

\begin{figure}
    \includegraphics[width=\hsize]{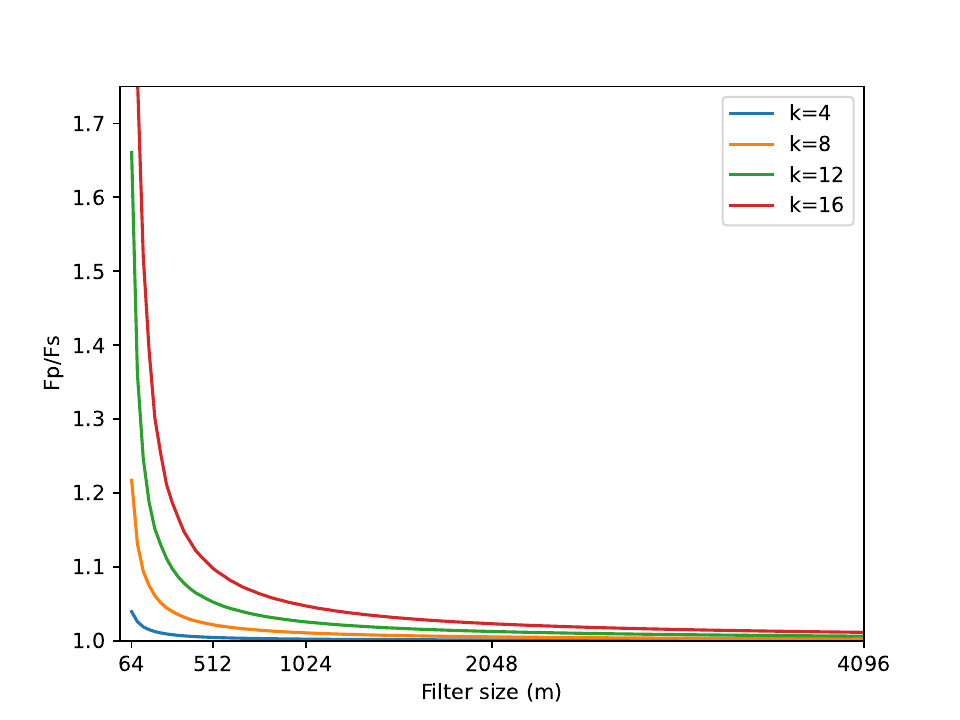}
  \caption{Ratio $F_p / F_s$ between partitioned and standard Bloom filters FPRs,
  for filters at nominal occupation ($n = \frac mk \ln 2$).}
\label{fig:std-vs-part}
\end{figure}

Figure~\ref{fig:std-vs-part} plots the ratio of false positives $F_p / F_s$
over $m$, for some values of $k$. For filters of cache line size (512 bits) or
larger, the difference between partitioned and standard filters is practically
irrelevant, and only for cases that are not practically usable due to
  minuscule capacity (very small high accuracy filters) would the
  difference be significant.

Table~\ref{tab:std-part-fp-comparison-occupations} shows the ratio of false
positives $F_p / F_s$ for filters at different occupations (namely $1/4$,
$1/2$, and $1/1$) relative to the nominal capacity. The ratio increases
somewhat for word sized filters and small occupations, but those occupations
for those filters are degenerate cases, with just a few elements inserted, and
negligible FPR, whether for standard or partitioned filters.

\begin{table}
  \caption{Ratio between partitioned and standard Bloom filters FPR, $F_p / F_s$, for different combinations of $m$, $k$, and
  occupation (fraction of the nominal capacity $n = \frac mk \ln 2$).}
  \label{tab:std-part-fp-comparison-occupations}
  \begin{center}
    \begin{tabular}{@{}rrrrr@{}}
  \toprule
  & & \multicolumn3c{occupation} \\
   \cmidrule(l){3-5}
  $m$ & $k$ & 1/4 & 1/2 & 1/1 \\
  \midrule
  \multirow2*{64}
  & 4 & 4.42059447 & 1.11720759 & 1.03941360 \\
  & 8 & 8.91227883 & 1.77601565 & 1.21703762 \\
  \midrule
  \multirow3*{512}
  & 4 & 1.00616297 & 1.00527287 & 1.00458411 \\
  & 8 & 1.02790590 & 1.02512045 & 1.02172097 \\
  & 16 & 1.13309283 & 1.11474124 & 1.09783475 \\
  \midrule
  \multirow3*{4096}
  & 4 & 1.00069246 & 1.00064963 & 1.00056676 \\
  & 8 & 1.00324437 & 1.00303940 & 1.00265094 \\
  & 16 & 1.01404449 & 1.01313731 & 1.01143019 \\
  \bottomrule
    \end{tabular}
  \end{center}
\end{table}

So, the average FPR is not relevant in practice for making a choice
between standard versus partitioned Bloom filters. But as we discuss next, a more
relevant issue is the distribution of false positives over the elements in
the domain subject to being tested.

\section{Weak spots in the domain}

There are two ways that Bloom filters can be used, and two different points of
view regarding false positives:
\begin{enumerate}
  \item Filter point of view: having a filter, in which elements were inserted
    along time, test new elements using the filter.
  \item Element point of view: for a specific element, test it against many
    different filters that show up, to see if the element is present in them.
\end{enumerate}

The first usage is the more normal, for which we want to know the global
average FPR. The second usage corresponds to the packet forwarding
scenario, where at each node (representing an element) many different filters
arrive (each one representing a path that a packet took to reach the node).
For this second usage we want to know, for each specific element in the
domain, the average FPR over all possible filters that do not include the element,
for each given combination of $k$, $m$, and $n$.
Particularly relevant is the question of whether this per-element
rate is the same for all elements (the global average) or whether it
is non-uniform, varying for different elements.

For partitioned Bloom filters, with $k$ independent parts, accessed by $k$
independent hash functions, the per-element FPR is the
same for all elements, and equal to the global average. But for standard Bloom
filters, the possibility of hash collisions makes some elements have less
than $k$ independent bits to test. We have thus a non-uniform distribution of false
positives: for a given element having $d < k$ different bit positions to test,
the average FPR will be higher than for those elements for
which no collisions occur. Elements suffering collisions are then
\emph{weak spots} in the domain: they will be considered more often than
expected as belonging to filters against which they are tested. As we will
see, for elements suffering several hash function collisions, the false
positive rate can be more than one order of magnitude larger than expected. We
now derive an exact formula for the per-element FPR.

\subsection{Per-element false positive analysis}

Consider a specific element $e$ of the domain, having $d$ different bit positions
resulting from the $k$ independent hash functions, where $d \leq k$. We want to
know the average FPR $F_s(n, m, k, d)$ when $e$ is tested
against standard Bloom filters of size $m$ where a set of $n$ elements not
containing $e$ was inserted.

A first observation is that the per-element rate cannot be obtained by simply
going to the exact formula in Equation~\ref{eq:fps}, where the fill ratio is
raised to the power of $k$, and replacing $(i / m)^k$ with $(i / m)^d$, i.e.,

\begin{equation}
  F_s(n, m, k, d) \neq \sum_{i=1}^m S(n, m, k, i) \left({i \over m}\right)^d.
\end{equation}

The reason is that by saying that there are $d$ different positions, they are
not independent, and we cannot use the independent testing assumption as for
the $k$ positions.
This can be seen by a simple example of a filter with $k=2$, $m=2$, $n=1$, and
computing the FPR for elements with $d=2$ different bits.
When considering the case $i=1$, i.e., one bit set in the filter, being the
fill ratio $1/2$, for $d=2$ there is no possibility of a false positive, while
using $(i/m)^d$ would give the erroneous $(1/2)^2 = 1/4$.

The correct formula for the probability of $d$ different bits being set when
$i$ of the $m$ bits in the filter are set is:
\begin{equation}
  \prod_{j=0}^{d-1} \frac{i-j}{m-j},
\end{equation}
i.e., the first of the $d$ positions is one of the $i$ bits set, the second
is one of the remaining $i-1$, the third one of the remaining $i-2$ and so on.
The probability is zero for $d > i$.

The correct formula for the per-element FPR is then obtained
by averaging over the different possible numbers of bits set, weighted by their
probability of occurring, as before, resulting in:

\begin{equation}
  F_s(n, m, k, d) = \sum_{i=1}^m S(n, m, k, i) \prod_{j=0}^{d-1}
  \frac{i-j}{m-j}.
\end{equation}

Table~\ref{tab:std-per-element-fp} shows how the per-element FPR compares with
the (global) average FPR, showing the ratio
$F_s(n, m, k, d) / F_s(n, m, k)$ for different numbers of hash collision $c =
k - d$, from no collision ($d=k$) up to three collisions ($d=k-3$), for
filters at different occupations (ratios relative to nominal capacity $n =
\frac mk \ln 2$).

\begin{table}
  \caption{Ratio between per-element and global FPR for
    standard Bloom filters, $F_s(n, m, k, d) / F_s(n, m, k)$, for different
    combinations of $m$, $k$, and hash collisions $c=k-d$, for filters at
  different occupations.}
  \label{tab:std-per-element-fp}
  \begin{center}
    \begin{tabular}{@{}crrrrrr@{}}
  \toprule
  & & & \multicolumn{4}c{collisions} \\
   \cmidrule(l){4-7}
  occupation & $m$ & $k$ & 0 & 1 & 2 & 3 \\
  \midrule
  \multirow4*{1/1}
                   & \multirow2*{64}
  & 4 & 0.91 & 1.88 & 3.85 & 7.78 \\
 && 8 & 0.59 & 1.39 & 3.25 & 7.47 \\
   \cmidrule(l){2-7}
                   & \multirow3*{512}
  & 8 & 0.95 & 1.92 & 3.89 & 7.88 \\
 && 16 & 0.79 & 1.62 & 3.31 & 6.78 \\
  \midrule
  \multirow4*{1/2}
                   & \multirow2*{64}
  & 4 & 0.76 & 3.21 & 12.92 & 50.00 \\
 && 8 & 0.14 & 1.09 & 7.50 & 47.38 \\
   \cmidrule(l){2-7}
                   & \multirow3*{512}
  & 8 & 0.87 & 3.09 & 10.87 & 38.10 \\
 && 16 & 0.56 & 2.03 & 7.41 & 26.86 \\
  \midrule
  \multirow4*{1/4}
                   & \multirow2*{64}
  & 4 & 0.38 & 5.90 & 74.86 & 804.25 \\
 && 8 & 0.00 & 0.07 & 4.45 & 134.33 \\
   \cmidrule(l){2-7}
                   & \multirow3*{512}
  & 8 & 0.74 & 5.03 & 33.76 & 224.42 \\
 && 16 & 0.23 & 1.87 & 15.28 & 123.08 \\
  \bottomrule
    \end{tabular}
  \end{center}
\end{table}

It can be seen that the FPR increases noticeably with
the number of hash collisions that occur for the element being tested, in
relation to the global average rate for the filter.
This effect is more prevalent for small occupations, with the FPR reaching two orders of magnitude larger than the global average for
$1/4$ occupation and three collisions.
This may cause surprises in scenarios where a filter is dimensioned with some
expectations about the FPR over its lifetime, from empty to
full. Some elements will incur much more false positives than what planned
for, if using either the standard or exact formula for the global average.


\subsection{Probability distribution of hash collisions}

The question of how frequent are those weak spots in the domain, specially the
``very weak'' spots having more than one hash collision is easily answered.
The probability of an element being a weak spot is an instance of the birthday
problem, as discussed above, with value given by Equation~\ref{eq:birthday}.
For an $m$ sized Bloom filter, the probability of the $k$ hashes resulting in
$d$ different bits (i.e., $c=k-d$ collisions) is an instance of the balls into
bins experiment, with value $B(k, m, d)$ as given by Equation~\ref{eq:balls-bins}.

Table~\ref{tab:collision-distribution} shows the probability of having some
(one or more) hash collisions, and of having exactly $0 \leq c \leq 3$ collisions,
for some combinations of $k$ and $m$.

\begin{table}
  \caption{Probability of having some hash collision(s) and of having exactly $0
    \leq c \leq 3$ hash collisions, for some combinations of $k$ and $m$.}
  \label{tab:collision-distribution}
  \begin{center}
    \begin{tabular}{@{}rrccccc@{}}
  \toprule
  & & \multicolumn{5}c{collisions} \\
   \cmidrule(l){3-7}
      $m$ &  $k$ & some & 0 & 1 & 2 & 3 \\
  \midrule
  \multirow2*{64}
  & 4 & 0.0911 & 0.9089 & 0.0894 & 0.0017 & 0.0000 \\
  & 8 & 0.3660 & 0.6340 & 0.3115 & 0.0510 & 0.0034 \\
  \midrule
  \multirow2*{512}
  & 8 & 0.0535 & 0.9465 & 0.0525 & 0.0010 & 0.0000 \\
  & 16 & 0.2108 & 0.7892 & 0.1905 & 0.0192 & 0.0011 \\
  \bottomrule
    \end{tabular}
  \end{center}
\end{table}

It can be seen that collisions happen frequently not only in word sized
filters (36\% of elements for $m=64$ and $k=8$) but also for the important
case of cache line sized blocks ($m=512$) in blocked Bloom filters, reaching
21\% for very high accuracy ($k=16$) filters. Two collisions can happen with
non-negligible frequency, in 5 percent of elements for the word sized filters
with $k=8$, or in two percent of elements in the ($m=512$, $k=16$) case.
And while three collisions is indeed very rare, 3 in a thousand for the ($m=64$,
$k=8$) filter or one in a thousand for the ($m=512$, $k=16$) filter, this is
no consolation when those ``unlucky'' elements are subject to being tested
against many filters.

\begin{figure}
    \includegraphics[width=\hsize]{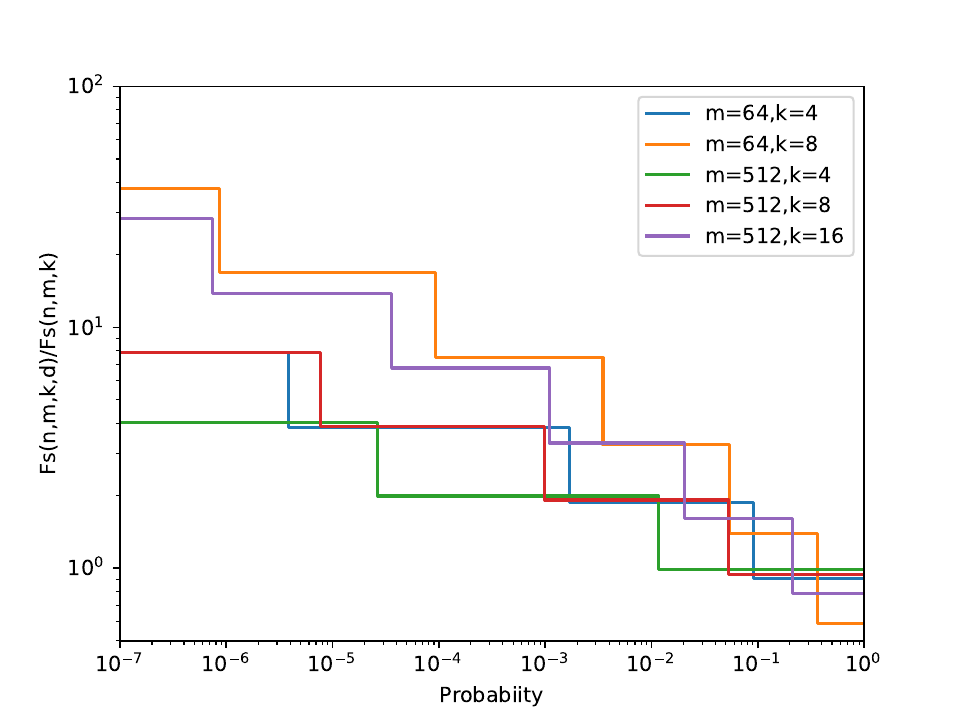}
  \caption{Inverse of tail distribution of ratio between per-element and global FPR for
    standard Bloom filters, $F_s(n, m, k, d) / F_s(n, m, k)$, for different
    combinations of $m$, $k$, for filters at nominal occupation ($n = \frac mk
    \ln 2$).}
\label{fig:tail-distributions}
\end{figure}

We now combine the per-element FPR values (as shown in
Table~\ref{tab:std-per-element-fp}) with the probabilities of their
occurrence (as shown in Table~\ref{tab:collision-distribution}.
Figure~\ref{fig:tail-distributions} plots the inverse of the tail distribution
  (ignoring values with very low probability of occurrence) of the
ratio between per-element and global FPR for standard Bloom filters, $F_s(n,
m, k, d) / F_s(n, m, k)$, for different combinations of $m$ and $k$, for
filters at nominal occupation ($n = \frac mk \ln 2$).
From it, it can be seen that, e.g., for a $m=512$, $k=16$ high accuracy
filter, for one in one thousand elements the FPR will be more than 6 times
the global FPR, and for almost one in a million elements it will be almost 30
times the global FPR.

\section{Pitfalls in double hashing}

One technique used to improve performance, by avoiding the need to compute $k$ hash
functions, is to resort to \emph{double hashing}, which amounts to using two
hash functions $\{h_1, h_2\}$, to simulate $k$ hash functions. In the more
naive form it amounts to computing $g_0, \ldots, g_{k-1}$ as:
\[
  g_i(x) = h_1(x) + i h_2(x)  \mod m
\]

The first time that double hashing was applied to Bloom filters seems to have
been by Dillinger and Manolios~\cite{DillingerM2004}, for model checking.
It was popularized after Mitzenmacher~\cite{KM2008} showed that it could be
used to implement a Bloom filter without any loss in the asymptotic false
positive probability, and experimentally validating it for medium sized Bloom
filters, starting with $m=10000$ bits.
However, small Bloom filters were not considered (e.g., a 512 bits block in a
BBF) and, as usual, only the global FPR was considered.

Here we address small filters and the possibility of a non-uniform
distribution of false positives, with weak spots in the domain. We show that
standard Bloom filters, but not partitioned ones, are prone to even more
problematic weak spots caused by the use of double hashing.  Although more
sophisticated variants, like enhanced double hashing or triple hashing have
been proposed, naive doubling hashing in particular has become relatively
popular, and can be found in many Bloom filter implementations.  Therefore,
these issues have practical consequences.

Dillinger's PhD dissertation~\cite{dillinger2010}, which includes a detailed
study of different forms of double and triple hashing, already recognized the
existence of pitfalls, specially in naive double hashing. It identified three
issues, which we now show that only affect standard, but not partitioned,
Bloom filters.

\begin{figure}
  \begin{center}
    \includegraphics[width=\hsize]{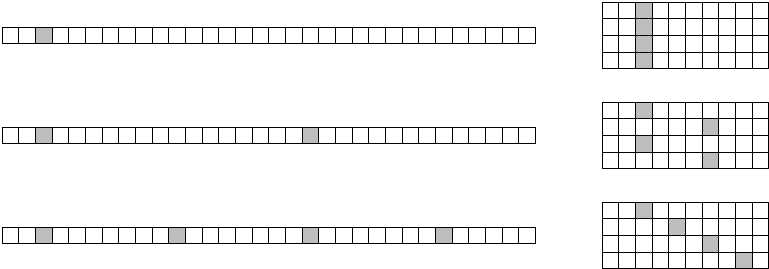}
  \end{center}
  \caption{Effects of double hashing when inserting an element $x$ in a
    standard (left) versus partitioned (right) Bloom filter, when $b=h_2(x)$
    is 0, $1/2$, or $1/4$ the size of the vector being indexed (filter or
  part).}
\label{fig:issue1}
\end{figure}

\paragraph{Issue 1}
Some possibilities for $b=h_2(x)$ can result in many repetitions of the same
index. The worse case would be if $b=0$ (mod $m$), in which case all indices
would be the same, but the existence of common factors between $b$ and $m$
also causes problems. Figure~\ref{fig:issue1} shows some examples, with $b=0$,
$b=m/2$ and $b=m/4$. On the left, for standard Bloom filters, there is
overwhelming index collision, which causes bit collisions, resulting in very
weak spots. In a BBF with 512 bit blocks, one out of 512 elements in the
domain will have a single bit set/tested, resulting in a disastrous $1/2$
probability of them being tested as a false positive in filters at nominal
capacity ($1/2$ fill ratio). Then, one out 512 elements $1/4$ probability, and so
on. For partitioned Bloom filters, index collisions do not
cause bit collisions, resulting always in $k$ bits being set/tested.

\begin{figure}
  \begin{center}
    \includegraphics[width=\hsize]{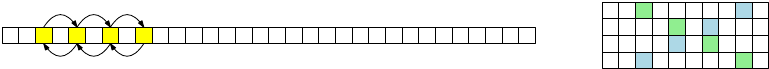}
  \end{center}
  \caption{Full overlap between $x$ and $y$ when using double hashing in a
  standard Bloom filter, when $h_1(y) = h_1(x) + (k-1) h_2(x) \mod m$ and 
  $h_2(y) = m - h_2(x)  \mod m$ (left), and the lack of such overlap in a
partitioned Bloom filter (right).}
\label{fig:issue2}
\end{figure}

\paragraph{Issue 2}
The indices generated by double hashing, used to index a standard Bloom filter
are treated as a set, not a sequence, and we can compute the same set going
``forward'' or going ``backward''. Two elements $x$ and $y$, can have a
full overlap of the $k$ bits without both $h_1$ and $h_2$ colliding, if
$h_1(y) = h_1(x) + (k-1) h_2(x) \mod m$ and 
$h_2(y) = m - h_2(x)  \mod m$. For a partitioned Bloom filter, such overlap
does not occur, as the different parts are indexed in order, and so we have
effectively a sequence of indices. Figure~\ref{fig:issue2} illustrates the
full overlap between $x$ and $y$, for a standard Bloom filter and the
absence of overlap in a partitioned Bloom filter.

\begin{figure}
  \begin{center}
    \includegraphics[width=\hsize]{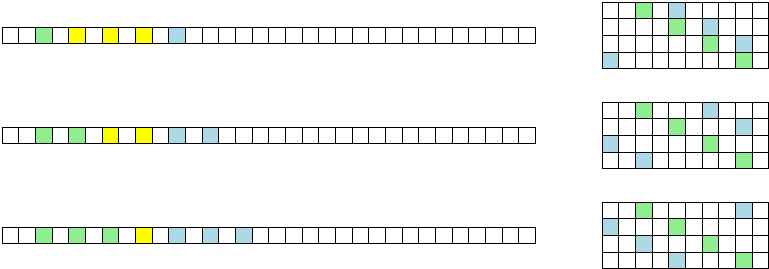}
  \end{center}
  \caption{Partial overlap (yellow) between $x$ (green) and $y$ (blue) when
    using double hashing in a standard Bloom filter, when $h_2(x) = h_2(y)
    \mod m$ (left), and the lack of such overlap in a partitioned Bloom filter (right).}
\label{fig:issue3}
\end{figure}

\paragraph{Issue 3}
Using double hashing in a standard Bloom filter is prone to partial
overlapping of the $k$ indices, namely when $h_2(x) = h_2(y) \mod m$. This is
illustrated in Figure~\ref{fig:issue3}. In the same figure, it can be seen
that in partitioned Bloom filters such overlap does not occur.

\medskip

Standard Bloom filters are thus subject to these anomalies, the more serious
being the possibility of extreme weak spots, if naive double hashing is used.
In theory, Issue 1 (which  causes weak spots) is easy to overcome, by ensuring
there are no collisions, e.g., in the popular case when $m$ is a power of two
by restricting $b=h_2(x)$ to produce odd numbers. In practice, implementers
have been sold the idea that double hashing can be used harmlessly, and
commonly do not take precautions, namely when the filter is parameterized,
being $m$ arbitrary and possibly small. This has occurred even in mainstream
libraries, such as in Google Core Libraries for Java~\cite{GuavaBF}.
Partitioned Bloom filters have the advantage of not being subject to such weak
spots, and thus are robust to naive double hashing implementations.

It should be noted that if Issue 1 is addressed, the impact of double hashing
on the global FPR is larger for partitioned Bloom filters than
for the standard ones. This impact comes from the probability of the pair of
indices for one element colliding with the pair from another element, i.e.,
$h_1(x) = h_1(y)$ and $h_2(x) = h_2(y)$ (modulo vector size). Between two elements
it is $1 / m^2$ for standard Bloom filters and $1 / (m/k)^2$ for partitioned.

In practice, for large Bloom filters the contribution of double hashing for
the global FPR is negligible, unless high accuracy filters are
wanted, in which case care must be taken and triple hashing may be needed.
For small filters, or in general when BBFs are used, neither double nor triple hashing
should be used, as only a few bits per index are needed, and a single hash
word can be split to obtain the $k$ indices. Concretely, in a BBF with 512 bit
blocks and $k=8$, we need 9 bits per index for standard and 6 bits per
index for partitioned filters. This means that a partitioned scheme needs
$6*8 = 48$ bits per block and a single 64 bit hash word is enough for filters
up to $2^{64-48} = 65536$ blocks, i.e., $2^{25} = 33554432$ bits, while if
standard filters are used $9*8 = 72$ bits per block are needed and a 64 bit
hash word is not enough even for small filters. This reinforces the
superiority of partitioned Bloom filters over standard ones.

\section{The flexibility advantage of disjoint parts}

Regardless of the FPR itself, the disjointness of the parts in
a partitioned Bloom filter provides several advantages over standard filters,
either in terms of obtaining fast implementations or making the partitioned
scheme more flexible to be used in more scenarios, or as the base for further
extensions. Each disjoint part can be sampled, extracted, added, or retired
individually, leading to interesting outcomes. We conclude our case by
surveying some of these usages and advantages.

\subsection{Fast Bloom filters through SIMD}

In addition to improving memory accesses, through blocked Bloom filters,
another way to improve performance is to use Single Instruction Multiple Data
(SIMD) processor extensions, to test multiple bits in a single processor cycle.
However, standard Bloom filters are not directly suitable to SIMD, because the
$k$ bits are spread over memory,
needing an extra \emph{gather} step to collect and place them appropriately,
causing some slowdown.

A sophisticated SIMD approach~\cite{PolychroniouR14} for standard
Bloom filters uses precisely \emph{gather} instructions to collect 
bits spread over memory. It achieves higher throughput, by testing different
hashes of different elements at each step, but not lower latency of individual
query operations.

Even using BBFs based on standard Bloom filter blocks is not directly
suitable to SIMD, because the $k$ bits are not placed over independent disjoint
parts of the cache line (e.g., words) to be used together as a vector register.
When introducing BBFs the authors already discussed SIMD usage, and to
overcome this problem they propose using a table of $k$ bit block-sized
patterns. However, to avoid collisions between elements when indexing, the table
cannot be too small, competing for cache usage.

Partitioned Bloom filters are more directly suitable to SIMD. A blocked Bloom
filter using the partitioned scheme, with cache-line sized blocks and word
sized parts is perfect for SIMD, and arises as the natural combination
of blocking and partitioning. This is precisely what \emph{Ultra-Fast Bloom
Filters}~\cite{LuWLZDWZL19} have recently proposed. We may conjecture that, had
partitioned Bloom filters been the norm at the time when BBFs were introduced,
this combination could have appeared one decade earlier.

\subsection{Set disjointness}

Bloom filters can also be used for set union and intersection. Unlike for
union (bitwise or) which is exact, intersection of filters (bitwise and)
over-represents the filter for the intersection: given sets $S_1$ and $S_2$,
we have $F(S_1) \land F(S_2) \geq F(S_1 \cap S_2)$.
In addition to testing for the presence of some element, an important use case
is testing for set disjointness, i.e., that the intersection is an empty set.
An example is checking whether two sets of addresses, representing a read-set
and a write-set are disjoint, when implementing \emph{transactional memory}.

Using standard Bloom filters, being sure that the sets are disjoint is only
possible when the resulting filter intersection is completely empty (all
zeroes). Having less than $k$ one bits is not enough, due to weak spots. As
already noticed~\cite{JeffreyS11}, even if the intersection result had a
single bit it could be (even if extremely unlikely) due to an element, present
in both sets, having the $k$ hash functions collide.

Partitioned Bloom filters are much better suited for testing set disjointness,
as it is enough that one of the $k$ parts of the filter intersection is empty
to conclude that the set intersection is empty. This was already
exploited~\cite{ceze2006bulk} for \emph{speculative multithreading}. A comparison
of set disjoitness testing concluded~\cite{JeffreyS11} that
the probability of \emph{false set-overlap} reporting was substantially
smaller for partitioned Bloom filters than standard Bloom filters. This
probability, for standard ($P_s$) and partitioned ($P_p$) $m$ sized filters
with $k$ hash functions, representing sets with $n_1$ and $n_2$ elements,
compares as:
\begin{align*}
  P_s & = 1 - \left(1 - {1 \over m}\right)^{k^2 n_1 n_2} \\
      & > 1 - \left(1 - {k \over m}\right)^{n_1 n_2} \\
      & > \left(1 - \left(1 - {k \over m}\right)^{n_1 n_2}\right)^k = P_p.
\end{align*}

This is intuitively easy to understand: the probability of a false set-overlap
for a standard $m$ sized filter, due to some of the $k*n_1*k*n_2$ pairs of indices
colliding, is greater than the probability of such an overlap in a given
$m/k$ sized part for the partitioned scheme, which is substantially greater
than the probability that there is an overlap in each of the $k$ parts.

\subsection{Size reduction}

Sometimes it is useful to obtain a smaller sized, lower accuracy, version of a
Bloom filter. Either because the filter was overdimensioned and we do not need
the resulting overly high accuracy; or we want to obtain an explicitly lower
accuracy view (but enough for some purpose), e.g., to ship over the network,
wanting to save bandwidth.

A standard Bloom filter is not suitable for this purpose because of the mingling
of bits from different hash functions. What can be done is to use the same $k$
hashes, but remap the indices to a smaller $m'$ sized vector (preferably with
$m$ some multiple of $m'$), moving the bit in position $i$ to
$i$ modulo $m'$, and using modulo $m'$ indexing for the new filter.
The problem is that the resulting fill rate renders the filter, when not
immediately useless, having an overly high FPR, when comparing
with the optimal for the new smaller size and the same number of
elements~\cite{PapapetrouSN10}.

\begin{table*}
  \caption{Comparison between Bloom filter variants and common fingerprint-based approaches.}
  \label{tab:feature-comparison}
  \begin{center}
    \begin{tabular}{@{}lcccccccc@{}}
  \toprule
      & \multicolumn4c{Bloom} & \multicolumn4c{fingerprint} \\
   \cmidrule(l){2-5} \cmidrule(l){6-9}
      & SBF & PBF & SBBF & PBBF & Cuckoo & Xor & Ribbon & BFF \\
  \midrule
      weak spots & yes & no & yes & no & no & no & no & no \\
      incremental construction & yes & yes & yes & yes & yes & no & no & no \\
      delete possible & no & no & no & no & yes & no & no & no \\
      construction time & normal & normal & fast & fast & slow & slow & normal--slowest & normal \\
      query time & normal & normal & fastest & fastest & fast & fast & normal & fast \\
      memory usage & high & high & highest & highest & normal & low & low-lowest & lower \\
      trivial intersection over-representation & high & low & high & low & - & - & - & - \\
      suitable for trivial disjointness test & no & yes & no & yes & - & - & - & - \\
      extract lower accuracy views & no & yes & no & yes & immut. & immut. & immut. & immut. \\
  \bottomrule
    \end{tabular}
  \end{center}
\end{table*}

Partitioned Bloom filters are much better for this purpose. Due to the
disjointness of the $k$ parts, we can simply consider the first $k'$
parts as a smaller Bloom filter, e.g., to be shipped elsewhere. For the worst
case of a filter already at full capacity, the new one will provide the
optimal FPR for the new smaller size. Considerable size reductions
are viable, which would render a standard Bloom filter useless due to the fill rate
approaching 1. The same paper proposes \emph{Block-partitioned
Bloom filters}, composed of several blocks (each block a standard filter, with
insertions in each block, and using AND for queries), to be able to extract some
blocks as a new filter. It mentions that maximum size flexibility is achieved by
using one hash per block, i.e., by using a partitioned Bloom filter.

\subsection{Duplicate detection in streams}

Bloom filter based approaches to achieve queries over a sliding window of an
infinite stream tend to be space inefficient. Traditionally they have been
based either on some variation of Counting Bloom filters~\cite{BonomiMPSV06},
on storing the insertion timestamp~\cite{ZhangG08}, or using several disjoint
segments which can be individually added and retired, one example being Double
Buffering~\cite{ChangLF04}. This uses a pair of \emph{active} and
\emph{warm-up} Bloom filters, using the active for queries and inserting in
both until the warm-up is half-full, at which point it becomes the active, the
previous active is discarded and a new empty warm-up is added.

While with standard Bloom filters a segment must be a whole filter,
partitioned Bloom filters can be used as a base for better designs, in which
each disjoint part can be treated as a segment. \emph{Age-Partitioned Bloom
Filters}~\cite{ShtulBA20} use $k + l$ (for some configurable $l$) parts in a
circular buffer, using the $k$ more ``recent'' parts for insertions,
discarding (zeroing) the ``oldest'' part after each generation (batch of
insertions), and testing for the presence of $k$ adjacent matches for queries.
This is the currently best Bloom filter based design for querying a
sliding window over a stream. As for the other usages, starting from
partitioned instead of standard filters was essential to be able to reach this
new design.

\section{Comparison}

Partitioned Bloom filters may have several feature advantages, compared with
  the standard, but is it enough for them to be competitive with alternative
  approaches? A common view is that Bloom filters have been superseded by
  fingerprint-based mechanisms, such as Cuckoo~\cite{FanAKM14} or
  Xor~\cite{GrafL20} filters. This is not necessarily the case: while that is
  true if the main concern is memory consumption and high accuracy, for
  moderate accuracy and when query time is important but memory less of a concern,
  Bloom filters, in blocked variants, remain the best~\cite{LangNKB19}.

We now summarize how the Bloom filter variants compare among themselves, and
  with some fingerprint-based approaches: the well known Cuckoo and Xor
  filters, and two recent state-of-the-art mechanisms, \emph{Ribbon
  filters}~\cite{DBLP:journals/corr/abs-2103-02515} and \emph{binary fuse
  filters} (BFF)~\cite{GrafL22}. Table~\ref{tab:feature-comparison} presents a
  feature-wise and qualitative comparison of filters (detailed quantitative
  results can be found
  elsewhere~\cite{GrafL20,DBLP:journals/corr/abs-2103-02515,GrafL22}). The
  columns, from left to right depict Bloom filters, in standard and
  partitioned variants, blocked Bloom filters using either standard or
  partitioned filters in blocks, and then fingerprint-based mechanisms: Cuckoo,
  Xor, Ribbon and binary fuse filters.

From all mechanisms, only standard Bloom filters (normal or blocked) suffer
from the weak-spots problem in the per-element FPR distribution.
Bloom approaches and cuckoo filters allow an incremental (progressive)
construction, starting from an empty set, essential for an ``online''
operation. The others, XOR-probe based filters, are immutable, having to be
built from a given set.
Cuckoo filters also allow deletes (Counting Bloom Filters~\cite{FCAB2000}
allow deletes but consume too much memory).

  Except for Ribbon, most fingerprint-based mechanisms are fast for queries,
  but lose to BBFs, which are also the fastest for construction. Cuckoo and
  Xor filters are slow to build.  Fingerprint-based mechanisms are better in
  memory usage, specially for high accuracy filters, in which case BBFs start
  to become too memory hungry, leading to the need for large blocks or
  multiblocking~\cite{PutzeSS09}.  If incremental construction is not needed,
  the two recent mechanisms are very appealing, BFF as very good both in
  memory usage and query speed, and Ribbon filters by being very configurable,
  allowing the lowest memory usage of all.

  Where partitioned Bloom filters shine (whether plain or blocked variants) is
  in allowing extra features, such as trivial intersections leading to less
  over-representation than standard Bloom filters, or being suitable for
  trivial disjoitness tests (e.g., for read/write sets in transactional
  memory). Cuckoo filters could allow such tests, but not in the trivial way
  Bloom approaches allow, 
  while the XOR-probe based filters do not allow intersections at all.
  Finally, partitioned Bloom filters allow extracting lower accuracy views,
  which can themselves be used as first class filters (allowing further
  insertions) while the fingerprint-based filters (including Cuckoo filters)
  only allow obtaining immutable views.

\section{Conclusions}

Frequently, a focus on one small difference in one quantitative aspect misses
the whole picture. Partitioned Bloom filters have thus been considered worse
than standard, and frequently not adopted, due to having slightly more false
positives. This is ironic given that the difference amounts to a negligible
variation of capacity, for the same FPR.

In this paper we have shown how much simpler, elegant, robust and versatile
partitioned Bloom filters are. The simplicity of the exact formula results from
the conceptual simplicity of them being essentially the AND of single-hash filters.
Standard Bloom filters have a more complex nature due to the possibility of
intra-element hash collisions, with a resulting complex exact formula,
normally approximated, leading sometimes to surprises.

But essentially, we have shown how standard Bloom filters exhibit a
non-uniform distribution of the false positive probability, with weak spots in
the domain: elements that are reported much more frequently as false positives than
expected. This is an aspect than has been neglected from the literature.
Moreover, the issue of weak spots is much aggravated when naive double hashing
is used. Even though easily circumventable, many libraries, including
mainstream ones, suffer from this anomaly. The lesson seems to be that
practitioners frequently skim over published results, failing to notice subtle
problems.
Partitioned Bloom filters have a uniform distribution of false positives over
the domain, with no weak spots, even if naive double hashing is used.
Moreover, the need for less hash bits makes such schemes less warranted.

Finally, going beyond set-membership test, by surveying other usages, the
flexibility of being able to sample, extract, add or retire individual parts
becomes clear, showing the partitioned scheme to be better.
Like the hardware community already did, partitioned Bloom filters should be
widely adopted by software implementers, and used as a better starting point
for new designs, replacing standard Bloom filters as the new normal.

\bibliographystyle{IEEEtran}
\bibliography{partitioned-bloom}

\begin{thebibliography}{10}
\providecommand{\url}[1]{#1}
\csname url@samestyle\endcsname
\providecommand{\newblock}{\relax}
\providecommand{\bibinfo}[2]{#2}
\providecommand{\BIBentrySTDinterwordspacing}{\spaceskip=0pt\relax}
\providecommand{\BIBentryALTinterwordstretchfactor}{4}
\providecommand{\BIBentryALTinterwordspacing}{\spaceskip=\fontdimen2\font plus
\BIBentryALTinterwordstretchfactor\fontdimen3\font minus
  \fontdimen4\font\relax}
\providecommand{\BIBforeignlanguage}[2]{{%
\expandafter\ifx\csname l@#1\endcsname\relax
\typeout{** WARNING: IEEEtran.bst: No hyphenation pattern has been}%
\typeout{** loaded for the language `#1'. Using the pattern for}%
\typeout{** the default language instead.}%
\else
\language=\csname l@#1\endcsname
\fi
#2}}
\providecommand{\BIBdecl}{\relax}
\BIBdecl

\bibitem{Bloom70}
B.~H. Bloom, ``Space/time trade-offs in hash coding with allowable errors,''
  \emph{Communications of the ACM}, vol.~13, no.~7, pp. 422--426, 1970.

\bibitem{broder2004network}
A.~Broder and M.~Mitzenmacher, ``Network applications of bloom filters: A
  survey,'' \emph{Internet mathematics}, vol.~1, no.~4, pp. 485--509, 2004.

\bibitem{tarkoma2012theory}
S.~Tarkoma, C.~E. Rothenberg, and E.~Lagerspetz, ``Theory and practice of bloom
  filters for distributed systems,'' \emph{IEEE Communications Surveys \&
  Tutorials}, vol.~14, no.~1, pp. 131--155, 2012.

\bibitem{Mullin1983}
J.~K. Mullin, ``A second look at bloom filters,'' \emph{Communications of the
  ACM}, vol.~26, no.~8, pp. 570--571, 1983.

\bibitem{ceze2006bulk}
L.~Ceze, J.~Tuck, J.~Torrellas, and C.~Cascaval, ``Bulk disambiguation of
  speculative threads in multiprocessors,'' \emph{ACM SIGARCH Computer
  Architecture News}, vol.~34, no.~2, pp. 227--238, 2006.

\bibitem{sanchez2007implementing}
D.~Sanchez, L.~Yen, M.~D. Hill, and K.~Sankaralingam, ``Implementing signatures
  for transactional memory,'' in \emph{40th Annual IEEE/ACM International
  Symposium on Microarchitecture (MICRO 2007)}.\hskip 1em plus 0.5em minus
  0.4em\relax IEEE, 2007, pp. 123--133.

\bibitem{dharmapurikar2003deep}
S.~Dharmapurikar, P.~Krishnamurthy, T.~Sproull, and J.~Lockwood, ``Deep packet
  inspection using parallel bloom filters,'' in \emph{High performance
  interconnects, 2003. proceedings. 11th symposium on}.\hskip 1em plus 0.5em
  minus 0.4em\relax IEEE, 2003, pp. 44--51.

\bibitem{KM2008}
A.~Kirsch and M.~Mitzenmacher, ``Less hashing, same performance: Building a
  better bloom filter,'' \emph{Random Struct. Algorithms}, vol.~33, no.~2, pp.
  187--218, 2008.

\bibitem{PutzeSS09}
\BIBentryALTinterwordspacing
F.~Putze, P.~Sanders, and J.~Singler, ``Cache-, hash-, and space-efficient
  bloom filters,'' \emph{{ACM} Journal of Experimental Algorithmics}, vol.~14,
  2009. [Online]. Available: \url{https://doi.org/10.1145/1498698.1594230}
\BIBentrySTDinterwordspacing

\bibitem{MRE1939}
{W. W. Rouse Ball. Revised by H. S. M. Coxeter}, \emph{Mathematical Recreations
  and Essays}, 11th~ed.\hskip 1em plus 0.5em minus 0.4em\relax Macmillan, 1939.

\bibitem{patblog2011}
P.~Blog, ``Who created the birthday problem, and even one more version,''
  \url{https://pballew.blogspot.com/2011/01/who-created-birthday-problem-and-even.html},
  2011 (accessed May 26, 2020).

\bibitem{mullin1987accessing}
J.~K. Mullin, ``Accessing textual documents using compressed indexes of arrays
  of small bloom filters,'' \emph{The Computer Journal}, vol.~30, no.~4, pp.
  343--348, 1987.

\bibitem{whitaker2002forwarding}
A.~Whitaker and D.~Wetherall, ``Forwarding without loops in icarus,'' in
  \emph{Open Architectures and Network Programming Proceedings, 2002
  IEEE}.\hskip 1em plus 0.5em minus 0.4em\relax IEEE, 2002, pp. 63--75.

\bibitem{LangNKB19}
\BIBentryALTinterwordspacing
H.~Lang, T.~Neumann, A.~Kemper, and P.~A. Boncz, ``Performance-optimal
  filtering: Bloom overtakes cuckoo at high-throughput,'' \emph{Proc. {VLDB}
  Endow.}, vol.~12, no.~5, pp. 502--515, 2019. [Online]. Available:
  \url{http://www.vldb.org/pvldb/vol12/p502-lang.pdf}
\BIBentrySTDinterwordspacing

\bibitem{FanAKM14}
\BIBentryALTinterwordspacing
B.~Fan, D.~G. Andersen, M.~Kaminsky, and M.~Mitzenmacher, ``Cuckoo filter:
  Practically better than bloom,'' in \emph{Proceedings of the 10th {ACM}
  International on Conference on emerging Networking Experiments and
  Technologies, CoNEXT 2014, Sydney, Australia, December 2-5, 2014}, 2014, pp.
  75--88. [Online]. Available: \url{https://doi.org/10.1145/2674005.2674994}
\BIBentrySTDinterwordspacing

\bibitem{BreslowJ18}
\BIBentryALTinterwordspacing
A.~Breslow and N.~Jayasena, ``Morton filters: Faster, space-efficient cuckoo
  filters via biasing, compression, and decoupled logical sparsity,''
  \emph{{PVLDB}}, vol.~11, no.~9, pp. 1041--1055, 2018. [Online]. Available:
  \url{http://www.vldb.org/pvldb/vol11/p1041-breslow.pdf}
\BIBentrySTDinterwordspacing

\bibitem{GrafL20}
\BIBentryALTinterwordspacing
T.~M. Graf and D.~Lemire, ``Xor filters,'' \emph{{ACM} J. Exp. Algorithmics},
  vol.~25, pp. 1--16, 2020. [Online]. Available:
  \url{https://doi.org/10.1145/3376122}
\BIBentrySTDinterwordspacing

\bibitem{manber1994algorithm}
U.~Manber and S.~Wu, ``An algorithm for approximate membership checking with
  application to password security,'' \emph{Information Processing Letters},
  vol.~50, no.~4, pp. 191--197, 1994.

\bibitem{Grandi2018}
\BIBentryALTinterwordspacing
F.~Grandi, ``On the analysis of bloom filters,'' \emph{Inf. Process. Lett.},
  vol. 129, pp. 35--39, 2018. [Online]. Available:
  \url{https://doi.org/10.1016/j.ipl.2017.09.004}
\BIBentrySTDinterwordspacing

\bibitem{Roberts79}
C.~S. Roberts, ``Partial-match retrieval via the method of superimposed
  codes,'' \emph{Proceedings of the {IEEE}}, vol.~67, no.~12, pp. 1624--1642,
  1979.

\bibitem{mitzenmacher2002compressed}
M.~Mitzenmacher, ``Compressed bloom filters,'' \emph{IEEE/ACM Transactions on
  Networking (TON)}, vol.~10, no.~5, pp. 604--612, 2002.

\bibitem{Bose2008210}
\BIBentryALTinterwordspacing
P.~Bose, H.~Guo, E.~Kranakis, A.~Maheshwari, P.~Morin, J.~Morrison, M.~Smid,
  and Y.~Tang, ``On the false-positive rate of bloom filters,''
  \emph{Information Processing Letters}, vol. 108, no.~4, pp. 210 -- 213, 2008.
  [Online]. Available:
  \url{http://www.sciencedirect.com/science/article/B6V0F-4SMWFJ6-1/2/5f479ecc8eea9b5b67869b97a54c16b6}
\BIBentrySTDinterwordspacing

\bibitem{christensen2010new}
K.~Christensen, A.~Roginsky, and M.~Jimeno, ``A new analysis of the false
  positive rate of a bloom filter,'' \emph{Information Processing Letters},
  vol. 110, no.~21, pp. 944--949, 2010.

\bibitem{Holder1889}
O.~H{\"o}lder, ``Ueber einen mittelwertsatz,'' \emph{Nachrichten von der
  Königl. Gesellschaft der Wissenschaften und der Georg-Augusts-Universität
  zu Göttingen}, no.~2, pp. 38--47, 1889.

\bibitem{jensen1906}
J.~L. W.~V. Jensen, ``Sur les fonctions convexes et les in{\'e}galit{\'e}s
  entre les valeurs moyennes,'' \emph{Acta mathematica}, vol.~30, pp. 175--193,
  1906.

\bibitem{gerrish1979}
F.~Gerrish, ``63.29. surjections from an m-set to an n-set,'' \emph{The
  Mathematical Gazette}, vol.~63, no. 426, pp. 259--261, 1979.

\bibitem{DillingerM2004}
\BIBentryALTinterwordspacing
P.~C. Dillinger and P.~Manolios, ``Fast and accurate bitstate verification for
  {SPIN},'' in \emph{Model Checking Software, 11th International {SPIN}
  Workshop, Barcelona, Spain, April 1-3, 2004, Proceedings}, ser. Lecture Notes
  in Computer Science, S.~Graf and L.~Mounier, Eds., vol. 2989.\hskip 1em plus
  0.5em minus 0.4em\relax Springer, 2004, pp. 57--75. [Online]. Available:
  \url{https://doi.org/10.1007/978-3-540-24732-6\_5}
\BIBentrySTDinterwordspacing

\bibitem{dillinger2010}
P.~C. Dillinger, ``Adaptive approximate state storage,'' Ph.D. dissertation,
  Northeastern University, 2010.

\bibitem{GuavaBF}
D.~Andreou and K.~A. Kluever, ``{BloomFilterStrategies} in google core
  libraries for java,''
  \url{https://github.com/google/guava/blob/master/guava/src/com/google/common/hash/BloomFilterStrategies.java},
  2011 (accessed September 8, 2020).

\bibitem{PolychroniouR14}
\BIBentryALTinterwordspacing
O.~Polychroniou and K.~A. Ross, ``Vectorized bloom filters for advanced {SIMD}
  processors,'' in \emph{Tenth International Workshop on Data Management on New
  Hardware, DaMoN 2014, Snowbird, UT, USA, June 23, 2014}, A.~Kemper and
  I.~Pandis, Eds.\hskip 1em plus 0.5em minus 0.4em\relax {ACM}, 2014, pp.
  6:1--6:6. [Online]. Available: \url{https://doi.org/10.1145/2619228.2619234}
\BIBentrySTDinterwordspacing

\bibitem{LuWLZDWZL19}
\BIBentryALTinterwordspacing
J.~Lu, Y.~Wan, Y.~Li, C.~Zhang, H.~Dai, Y.~Wang, G.~Zhang, and B.~Liu,
  ``Ultra-fast bloom filters using {SIMD} techniques,'' \emph{{IEEE} Trans.
  Parallel Distrib. Syst.}, vol.~30, no.~4, pp. 953--964, 2019. [Online].
  Available: \url{https://doi.org/10.1109/TPDS.2018.2869889}
\BIBentrySTDinterwordspacing

\bibitem{JeffreyS11}
\BIBentryALTinterwordspacing
M.~C. Jeffrey and J.~G. Steffan, ``Understanding bloom filter intersection for
  lazy address-set disambiguation,'' in \emph{{SPAA} 2011: Proceedings of the
  23rd Annual {ACM} Symposium on Parallelism in Algorithms and Architectures,
  San Jose, CA, USA, June 4-6, 2011 (Co-located with {FCRC} 2011)},
  R.~Rajaraman and F.~M. on~the Heath, Eds.\hskip 1em plus 0.5em minus
  0.4em\relax {ACM}, 2011, pp. 345--354. [Online]. Available:
  \url{https://doi.org/10.1145/1989493.1989551}
\BIBentrySTDinterwordspacing

\bibitem{PapapetrouSN10}
\BIBentryALTinterwordspacing
O.~Papapetrou, W.~Siberski, and W.~Nejdl, ``Cardinality estimation and dynamic
  length adaptation for bloom filters,'' \emph{Distributed Parallel Databases},
  vol.~28, no. 2-3, pp. 119--156, 2010. [Online]. Available:
  \url{https://doi.org/10.1007/s10619-010-7067-2}
\BIBentrySTDinterwordspacing

\bibitem{BonomiMPSV06}
\BIBentryALTinterwordspacing
F.~Bonomi, M.~Mitzenmacher, R.~Panigrahy, S.~Singh, and G.~Varghese, ``An
  improved construction for counting bloom filters,'' in \emph{Algorithms -
  {ESA} 2006, 14th Annual European Symposium, Zurich, Switzerland, September
  11-13, 2006, Proceedings}, 2006, pp. 684--695. [Online]. Available:
  \url{https://doi.org/10.1007/11841036\_61}
\BIBentrySTDinterwordspacing

\bibitem{ZhangG08}
\BIBentryALTinterwordspacing
L.~Zhang and Y.~Guan, ``Detecting click fraud in pay-per-click streams of
  online advertising networks,'' in \emph{28th {IEEE} International Conference
  on Distributed Computing Systems ({ICDCS} 2008), 17-20 June 2008, Beijing,
  China}, 2008, pp. 77--84. [Online]. Available:
  \url{https://doi.org/10.1109/ICDCS.2008.98}
\BIBentrySTDinterwordspacing

\bibitem{ChangLF04}
\BIBentryALTinterwordspacing
F.~Chang, K.~Li, and W.~Feng, ``Approximate caches for packet classification,''
  in \emph{Proceedings {IEEE} {INFOCOM} 2004, The 23rd Annual Joint Conference
  of the {IEEE} Computer and Communications Societies, Hong Kong, China, March
  7-11, 2004}, 2004, pp. 2196--2207. [Online]. Available:
  \url{https://doi.org/10.1109/INFCOM.2004.1354643}
\BIBentrySTDinterwordspacing

\bibitem{ShtulBA20}
\BIBentryALTinterwordspacing
A.~Shtul, C.~Baquero, and P.~S. Almeida, ``Age-partitioned bloom filters,''
  \emph{CoRR}, vol. abs/2001.03147, 2020. [Online]. Available:
  \url{http://arxiv.org/abs/2001.03147}
\BIBentrySTDinterwordspacing

\bibitem{DBLP:journals/corr/abs-2103-02515}
\BIBentryALTinterwordspacing
P.~C. Dillinger and S.~Walzer, ``Ribbon filter: practically smaller than bloom
  and xor,'' \emph{CoRR}, vol. abs/2103.02515, 2021. [Online]. Available:
  \url{https://arxiv.org/abs/2103.02515}
\BIBentrySTDinterwordspacing

\bibitem{GrafL22}
\BIBentryALTinterwordspacing
T.~M. Graf and D.~Lemire, ``Binary fuse filters: Fast and smaller than xor
  filters,'' \emph{{ACM} J. Exp. Algorithmics}, vol.~27, pp. 1.5:1--1.5:15,
  2022. [Online]. Available: \url{https://doi.org/10.1145/3510449}
\BIBentrySTDinterwordspacing

\bibitem{FCAB2000}
L.~Fan, P.~Cao, J.~Almeida, and A.~Z. Broder, ``Summary cache: a scalable
  wide-area web cache sharing protocol,'' \emph{IEEE/ACM Trans. Netw.}, vol.~8,
  no.~3, pp. 281--293, 2000.

\end{thebibliography}

\begin{IEEEbiography}
  [{\includegraphics[width=1in,height=1.25in,clip,keepaspectratio]{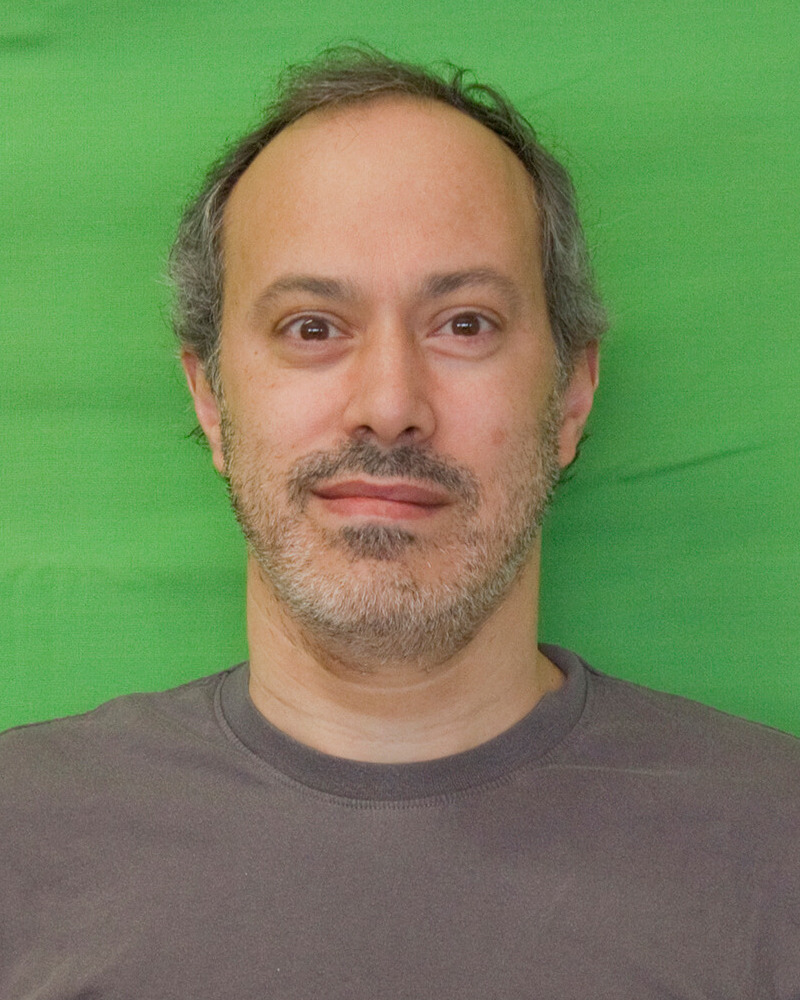}}]
  {Paulo S\'ergio Almeida}
is an assistant professor at University of Minho, and a senior researcher at
HASLab / INESC TEC. He obtained a MSc degree from University of Porto in 1994
and a PhD degree in Computer Science from Imperial College London in 1998.
His research activities have been in the area of distributed systems. Some main
contributions have been in causality tracking mechanisms, eventually
consistent databases, and fault-tolerant distributed aggregation
algorithms. Other subjects have been Bloom filters and distributed algorithms
in graphs. For some years the main focus of research has been on CRDTs
(Conflict-free Replicated Data Types), with some results being the development
  of both pure operation-based and delta-state based CRDTs.  Recently he has
been revisiting Bloom filters.
\end{IEEEbiography}

\end{document}